\documentclass[useAMS,usenatbib]{mn2e}
\usepackage{eso-pic,graphicx}

\title[A photometric and spectroscopic study of NSVS\,14256825]
  {A photometric and spectroscopic study of NSVS\,14256825: the second sdOB+dM eclipsing 
   binary\thanks{Based on observations carried out at the Observat\'orio do 
   Pico dos Dias (OPD/LNA), in Brazil}}

\author[L.~A.~Almeida et al.]
  {L.~A.~Almeida\thanks{E-mail: leonardo@das.inpe.br},
  F.~Jablonski, J.~Tello and C. V. Rodrigues            
   \\
  Instituto Nacional de Pesquisas Espaciais/MCTI, 
      Av. dos Astronautas 1758, S\~ao Jos\'e dos Campos, SP, 12227-010, Brazil\\
  }
\date{Released 2002 Xxxxx XX}

\pagerange{\pageref{firstpage}--\pageref{lastpage}} \pubyear{2011}

\def\LaTeX{L\kern-.36em\raise.3ex\hbox{a}\kern-.15em
    T\kern-.1667em\lower.7ex\hbox{E}\kern-.125emX}

\begin{document}

\label{firstpage}

\maketitle

\begin{abstract}
We present an analysis of UBVR$_{\rm C}$I$_{\rm C}$JH photometry and 
phase-resolved optical spectroscopy of NSVS\,14256825, an HW Vir type binary. 
The members of this class consist of a hot subdwarf and a main-sequence 
low-mass star in a close orbit ($P_{\rm orb} \sim 0.1$~d). Using the primary-eclipse 
timings, we refine the ephemeris for the system, which has an orbital period
of 0.11037 d. From the spectroscopic data analysis, we derive the 
effective temperature,  $T_1 = 40000 \pm 500$~K, the surface gravity, 
$\log g_1 = 5.50\pm0.05$, and the helium abundance, $n(\rm He)$/$n(\rm H)=0.003\pm0.001$,
for the hot component. Simultaneously modelling the photometric and spectroscopic data 
using the Wilson-Devinney code, we obtain the geometrical and physical parameters of 
NSVS\,14256825. 
Using the fitted orbital inclination and mass ratio ($i = 82\fdg5\pm0\fdg3$ and
$q = M_2/M_1 = 0.260\pm0.012$, respectively), the components of the system have 
$M_1 = 0.419 \pm 0.070~\rm M_{\odot}$, $R_1 = 0.188 \pm 0.010~\rm R_{\odot}$, 
$M_2 = 0.109 \pm 0.023~\rm M_{\odot}$, and $R_2 = 0.162 \pm 0.008~\rm R_{\odot}$.  
From its spectral characteristics, the hot star is classified as an sdOB star.
\end{abstract}


\section{Introduction}

HW Virginis (HW Vir) systems consist of a subdwarf B or OB (sdB or sdOB; hereafter 
referred to as sdB) plus a main sequence star, which form an eclipsing pair in a compact 
orbit, $P_{\rm orb} \sim 0.1$ d \citep{Heber2009}. These systems 
are believed to evolve through a common envelope phase when the primary (sdB) is a 
red giant. During this stage, the secondary star (dM) spirals in towards the primary one and the 
potential gravitational energy released is absorbed by the envelope, which is subsequently 
ejected \citep{Taam2010}. The final separation between the dM and the sdB stars 
depends on the initial mass ratio of the binary and the initial separation. 

sdB stars consist of a helium-burning core covered by a thin hydrogen-dominated envelope. 
The atmosphere abundance is normally $n$(He)$/n$(H) $\sim$ 0.01, the effective temperatures are in the range of 22000--37000 K, and the logarithmic surface gravities are normally between 5.2 and 5.7.
They populate a narrow strip on the extreme horizontal branch (EHB) in
the Hertzsprung-Russell diagram. The single-star stellar evolution predicts a narrow
mass range:  0.46 -- 0.50 \,M$_{\odot}$ (Dorman et al. 1993).  On the other hand, models
based on binary evolution predict a broader range of masses, from 0.3 to 0.8 \,M$_{\odot}$ 
(Han et al. 2003). Hence an important step in understanding the origin of sdB stars is the 
determination of their mass distribution. A recent review of sdB stars is 
presented by \citet{Heber2009}.

There are currently ten members of the HW Vir class, whose main features are summarised
in Table~\ref{tab:known}. Among them,  NSVS\,14256825 (2MASS\,J2020+0437; hereafter 
referred to as NSVS\,1425) is one of the least 
studied. It was discovered in the public data from the Northern Sky Variability 
Survey \citep{Wozniak2004}. The sole information on this system comes from the 
photometric data by \citet*{Wils2007}. These authors obtained B, V, and I$_{\rm C}$ 
photometric light curves. The main parameters obtained by \citet{Wils2007} are listed 
in Table~\ref{tab:known}.

\begin{table*}
\begin{center}
\label{tab:known}
\begin{minipage}{176mm}
\caption{The currently known sdB$+$dM eclipsing binaries.}
\begin{tabular}{llllllllll}
\hline
\hline
Name             & $T_{1}$ &$M_{1}$     &  $M_{2}$   & $R_{1}$    & $R_{2}$    &$\log g_1$& Period  &Refs.& Notes\\
                 &  (K)    &($\rm M_{\sun}$)&($\rm R_{\sun}$)&($\rm R_{\sun}$)&($\rm R_{\sun}$)&          & (d)     &     &         \\
\hline
AA Dor & 42000   & 0.33/0.47 &0.064/0.079&0.179/0.20 &0.097/0.108& 5.46& 0.261 &\footnotemark[1]$^,$\footnotemark[2]$^,$\footnotemark[3] & \footnotemark[15]$^,$\footnotemark[16] \\
                 &         &   0.25     & 0.054      &  0.165     &  0.089     &          &         &\footnotemark[4] & \footnotemark[16]  \\ 
NSVS\,14256825   & 40000   &   0.419    & 0.109      & 0.190      &  0.151     & 5.50     & 0.11037 & this work       & \footnotemark[15]$^,$\footnotemark[16] \\
                 & --      &   0.46     & 0.21       & --         & --         & --       & 0.11040 &\footnotemark[5] & \footnotemark[15] \\                 
PG 1336$-$018    & 31300   &   0.389    & 0.110      & 0.150      &  0.160     & 5.60     & 0.10102 &\footnotemark[6] & \footnotemark[15]$^,$\footnotemark[16] \\
                 & 32740   &   0.459    & --         & --         &  --        & --       & --      &\footnotemark[7] & \footnotemark[17] \\
2M 1533+3759     & 30400   &   0.377    & 0.113      & 0.166      &  0.152     & 5.58     & 0.16177 &\footnotemark[8] & \footnotemark[15]$^,$\footnotemark[16] \\
2M 1938$+$4603   & 29564   &   0.48     & 0.12       & --         &  --        & 5.425    & 0.12576 &\footnotemark[9] & \footnotemark[15]$^,$\footnotemark[16] \\
HS 0705+6700     & 28800   &   0.48     & 0.13       & 0.230      &  0.186     & 5.40     & 0.09565 &\footnotemark[10]& \footnotemark[15]$^,$\footnotemark[16] \\
PG 1241$-$084    & 28488   &   0.48     & 0.14       & 0.176      &  0.180     & 5.63     & 0.11676 &\footnotemark[11]& \footnotemark[15] \\
HS 2231+2441     & 28370   & 0.47/0.499 & 0.075/0.072& 0.250      &  0.127     & 5.39     & 0.11059 &\footnotemark[12]& \footnotemark[15]$^,$\footnotemark[16] \\
SDSSJ0820$+$0008 & 26700   & 0.25/0.47  & 0.068/0.045& --         &  --        & 5.48     & 0.096   &\footnotemark[13]& \footnotemark[15]$^,$\footnotemark[16] \\
BUL SC16\,335    & --      &   --       & --         & --         &  --        & --       & 0.12505 &\footnotemark[14]& \footnotemark[15] \\
\hline
\end{tabular}\\
\footnotemark[1]{\citet{Heber2009}};~
\footnotemark[2]{\citet{2011A&A...531L...7K}};~
\footnotemark[3]{\citet{Hilditch2003}};~
\footnotemark[4]{\citet{2009MNRAS.395.2299R}};~
\footnotemark[5]{\citet{Wils2007}};~
\footnotemark[6]{\citet{Vuckovic2009}};~
\footnotemark[7]{\citet{Charpinet2008}};~
\footnotemark[8]{\citet{For2010}};~
\footnotemark[9]{\citet{Ostensen2010}};~
\footnotemark[10]{\citet{Drechsel2001}};~
\footnotemark[11]{\citet{Wood1999}};~
\footnotemark[12]{\citet{Ostensen2007}};~
\footnotemark[13]{\citet{Geier2011}};~
\footnotemark[14]{\citet{Polubek2007}};~
\footnotemark[15]{Light Curves};~
\footnotemark[16]{Spectroscopy};~
\footnotemark[17]{Asteroseismology}.
\end{minipage}

\end{center}
\end{table*}

In this paper, we report on multi-band photometry and phase-resolved optical
spectroscopy of NSVS\,1425. We present an improved solution for its geometrical 
and physical parameters and discuss these results in the context of HW Vir systems
and their evolution.

\section{Observations and Data Reduction}

\subsection{Optical and near infrared photometry}

The observations were carried out using the Observat\'orio do Pico 
dos Dias (OPD/LNA) facilities, in Brazil. Photometric data in the U, B, V, R$_{\rm C}$, I$_{\rm C}$, J, and H bands were obtained from July to November, 2010. Optical observations 
were performed using a CCD camera attached to the 0.6-m IAG telescope. Near infrared 
photometry data were collected by means of a {\tt CamIV} imager attached to the 1.6-m 
Perkin-Elmer telescope. 

The procedure to remove undesired effects from the CCD data included obtaining 
$\sim$100 bias frames and $\sim$30 dome flat-field images for each night 
of observations. The NIR flat-field images were produced using separate sequences of $30$ 
{``on"} and $30$ {``off"} exposures. The resulting of the ``on"\ image minus ``off"\ image was 
used as the master flat-field image. Table~\ref{table:1} summarises the characteristics 
of the data collected for NSVS\,1425. In this table, $N$ is the number of individual 
images obtained with the integration time t$_{\rm exp}$.

\begin{figure}
 \resizebox{\hsize}{!}{\includegraphics[angle=270]{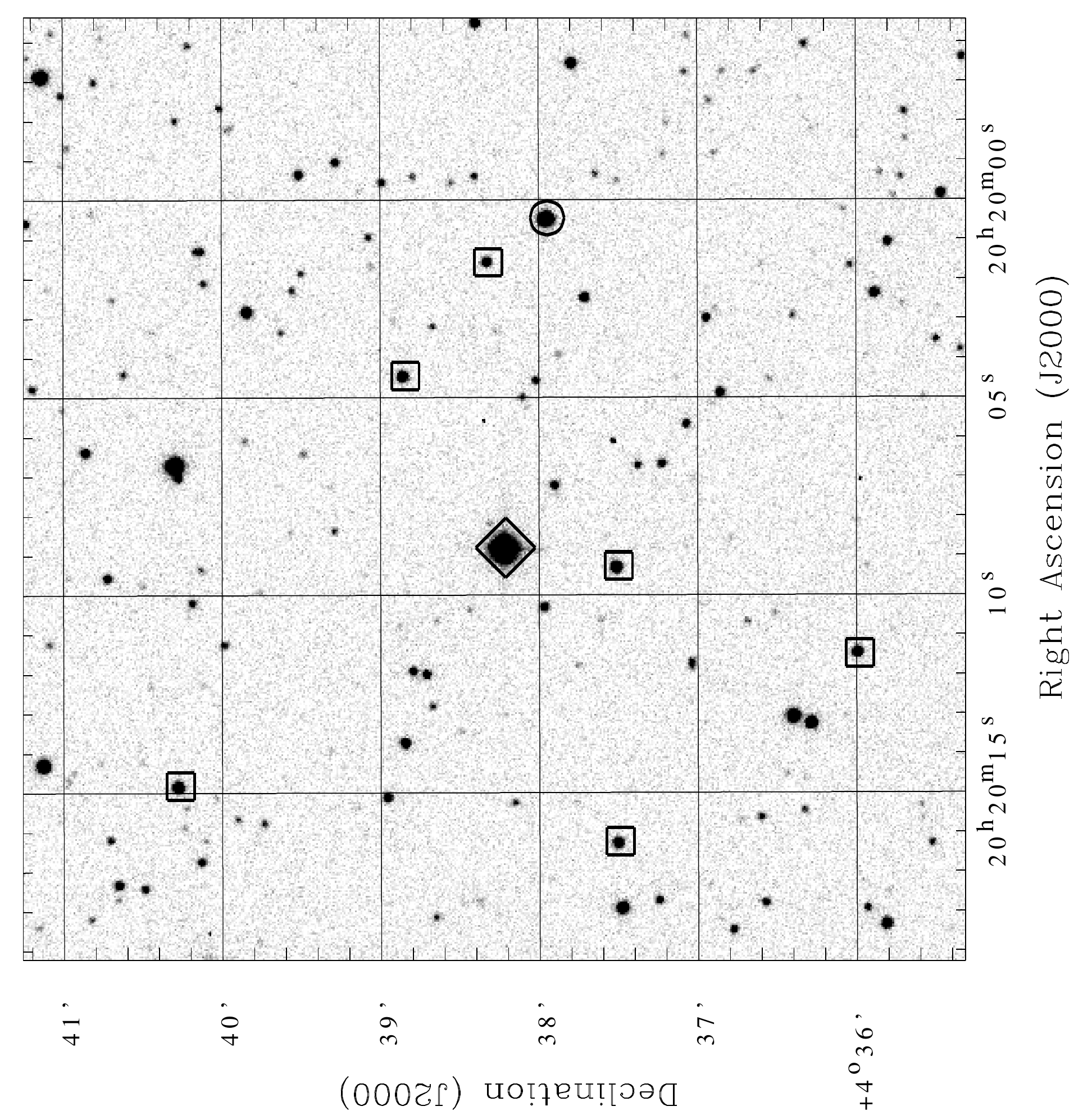}}
 \caption{Finding chart for NSVS\,14256825 in the R$_{\rm C}$ band obtained using the 
          OPD/LNA 0.6-m telescope. The circle shows NSVS\,1425, the diamond is 
          the adopted reference star, and the squares outline additional comparison stars.}
 \label{chart}
\end{figure}

\begin{figure}
 \resizebox{\hsize}{!}{\includegraphics[angle=-90]{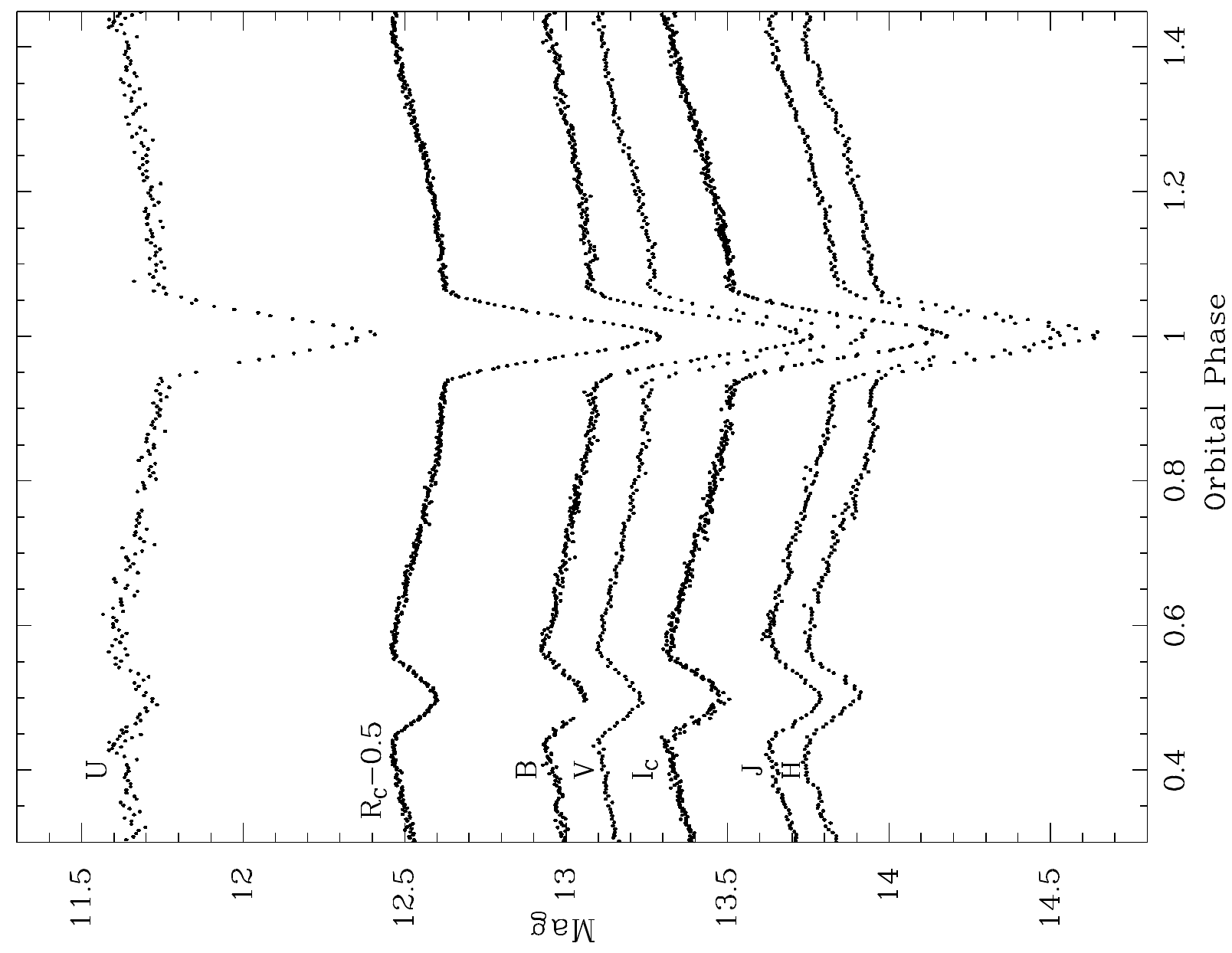}}
 \caption{Calibrated NSVS\,1425 light curves in the U, B, V, R$_{\rm C}$, 
          I$_{\rm C}$, J, and H bands folded on the 0.1104~d orbital period. The 
          R$_{\rm C}$-band curve was displaced upwards by 0.5 mag to improve visualisation.}
 \label{light_curve}
\end{figure}

The preparation of the CCD data was performed using standard {\tt IRAF}\footnote{http://www.iraf.noao.edu}
tasks \citep{Tody1993} and consisted of subtracting a master median bias image from 
each program image, and then dividing the result by a normalised flat-field frame. 
In the J and H bands, additional steps of linearization and sky subtraction from 
dithered images were used in the preparation of the data. For both optical and
infrared data, differential photometry was used to obtain the relative fluxes between 
the target and a set of constant flux stars in the field of view. As the NSVS\,1425 field is not crowded, the extraction of the fluxes was carried out using aperture photometry. 
Figure~\ref{chart} shows a finding chart for NSVS\,1425 with a reference star and a 
number of additional comparison stars adopted for differential photometry. Photometric 
standard stars were observed each night in order to calibrate the optical data. 
The NIR calibration is directly provided by the 2MASS catalogue magnitudes for the 
reference and comparison stars. Figure~\ref{light_curve} shows a sample of calibrated 
light curves folded on the NSVS\,1425 orbital period.  

As can be seen in Figure~\ref{light_curve}, the light curves of NSVS\,1425
show a prominent reflection effect, which increases towards longer wavelengths. 
The depth of the primary eclipse is $\sim$0.7\,mag and does not change significantly with 
wavelength, while the depth of the secondary eclipse increases
towards longer wavelengths, from $\sim$0.1\,mag in the U band to $\sim$0.18\,mag 
in the H band. Table~\ref{apparent_mag}~ shows the apparent magnitudes for 
NSVS\,1425 at primary and secondary minima.

\begin{table}
\caption{Log of the photometric observations.}           
\label{table:1} 
\centering                         
\begin{tabular}{l c c c c}        
\hline\hline                 
UT Date~~~~~ & $N$ & \ t$_{\rm exp}$(s) & Telescope & Filter  \\  
\hline                       
   2010 Jul 30  & 300 & 20 & 0.6-m & R$_{\rm C}$  \\
   2010 Jul 30  & 185 & 30 & 1.6-m & J  \\
   2010 Jul 31  & 250 & 30 & 0.6-m & B  \\
   2010 Jul 31  & 450 & 20 & 0.6-m & R$_{\rm C}$  \\
   2010 Jul 31  & 177 & 30 & 1.6-m & J  \\
   2010 Jul 31  & 235 & 30 & 1.6-m & H  \\
   2010 Aug 01  & 235 & 40 & 0.6-m & U  \\
   2010 Aug 01  & 75  & 40 & 0.6-m & B  \\
   2010 Aug 01  & 100 & 30 & 1.6-m & Y  \\
   2010 Aug 01  & 215 & 30 & 1.6-m & J  \\
   2010 Aug 02  & 230 & 40 & 0.6-m & U \\
   2010 Aug 02  & 400 & 20 & 0.6-m & I$_{\rm C}$ \\
   2010 Aug 06  & 120 & 30 & 0.6-m & V \\
   2010 Aug 07  & 220 & 30 & 0.6-m & V \\
   2010 Aug 08  & 220 & 30 & 0.6-m & I$_{\rm C}$ \\
   2010 Aug 09  & 160 & 25 & 0.6-m & R$_{\rm C}$ \\
   2010 Aug 10  &  80 & 25 & 0.6-m & R$_{\rm C}$ \\
   2010 Aug 18  & 800 & 10 & 0.6-m & R$_{\rm C}$ \\
   2010 Aug 18  & 530 & 15 & 0.6-m & I$_{\rm C}$ \\
   2010 Aug 19  & 225 & 40 & 0.6-m & U \\
   2010 Aug 20  & 420 & 20 & 0.6-m & B \\
   2010 Sep 01  & 160 & 30 & 0.6-m & V \\
   2010 Sep 03  & 250 & 30 & 1.6-m & J \\
   2010 Sep 04  & 142 & 50 & 1.6-m & J \\
   2010 Nov 03  & 352 & 20 & 0.6-m & I$_{\rm C}$ \\
\hline                                 
\end{tabular}
\end{table}

\subsection{Optical spectroscopy}

The spectroscopic observations were performed using the Cassegrain 
spectrograph attached to the 1.6-m telescope at OPD/LNA. Thirty six (36) spectra were obtained using the 1200 l/mm grating and integration times of 10 or 15 minutes. The spectral 
coverage of this configuration is 3950-4900 \AA, using 1.8~\AA~resolution 
(from the FWHM of the wavelength calibration lines).  A hundred (100) bias frames and 30 flat-field frames were obtained each night to remove systematic signatures from the CCD detector. Observations of a He-Ar comparison lamp were made every two exposures on the target to provide
wavelength calibration. The spectrophotometric standard stars HR~1544, HR~7596, and 
HR~9087 \citep{Hamuy1992} were observed for flux calibration. 

\begin{figure}
 \resizebox{\hsize}{!}{\includegraphics[angle=-90]{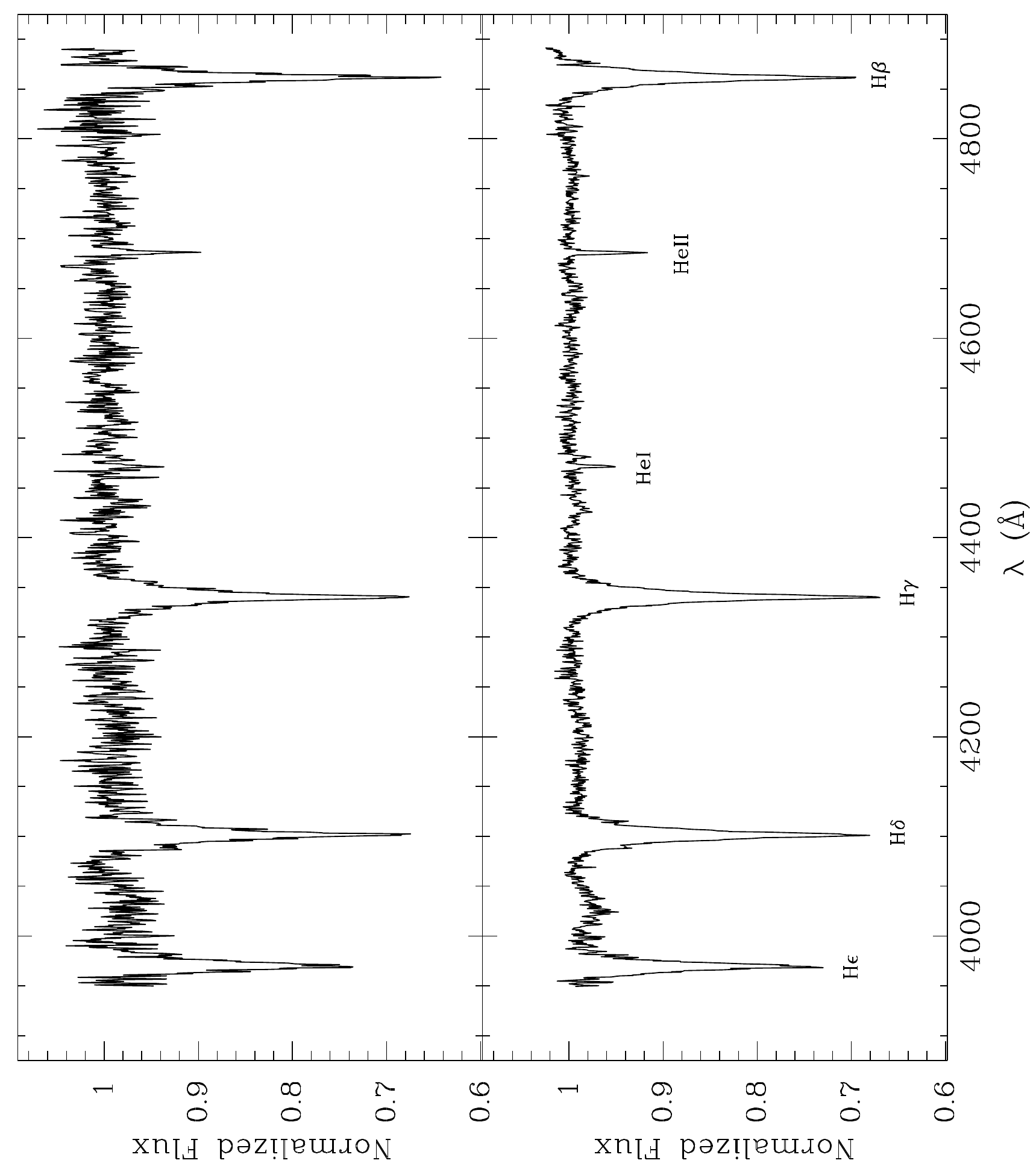}}
 \caption{Upper panel: a normalised individual spectrum of NSVS\,1425 with 
          900 s integration time.
          Lower panel: the average of 36 spectra after correcting for orbital 
          motion.}
 \label{fig:spectrum}
\end{figure}

The reduction of the spectra was carried out following the steps of bias subtraction, 
flat-field structure removal, optimal extraction, wavelength calibration, and flux 
calibration using the standard routines in {\tt IRAF}. The upper panel in 
Figure~\ref{fig:spectrum} shows a typical normalised individual spectrum. The lower 
panel shows the average of all spectra after Doppler shifting using the radial 
velocity orbital solution (see Section~\ref{sec:rvsolution}).

\section{Analysis and Results}

\begin{table}
\caption{Apparent magnitudes of NSVS\,1425 at primary and secondary minima.}   
\label{apparent_mag}      
\centering                          
\begin{tabular}{l r r r}       
\hline\hline               
 UT Date   & Band   &  Primary & Secondary \\
           &        &  minimum & minimum   \\
\hline                      
  2010 Aug 02 &U           & 12.41$\pm$0.20  & 11.73$\pm$0.20   \\
  2010 Aug 20 &B           & 13.76$\pm$0.15  & 13.06$\pm$0.15   \\
  2010 Aug 07 &V           & 13.93$\pm$0.13  & 13.23$\pm$0.13  \\
  2010 Jul 31 &R$_{\rm C}$ & 13.79$\pm$0.12  & 13.10$\pm$0.12  \\
  2010 Aug 18 &I$_{\rm C}$ & 14.18$\pm$0.16  & 13.47$\pm$0.16  \\
  2010 Aug 01 &J           & 14.50$\pm$0.24  & 13.80$\pm$0.24  \\
  2010 Jul 31 &H           & 14.60$\pm$0.25  & 13.91$\pm$0.25  \\
\hline
\end{tabular}
\end{table}

\subsection{Ephemeris}\label{ephemeris}

To determine an ephemeris for the times of the primary minimum in NSVS\,1425, we combined 
our timings and those from \citet{Wils2007}, after converting them to barycentric dynamical 
time (TDB). Our eclipse timings were obtained by modelling the primary eclipse using the 
Wilson-Devinney (WD) code \citep{Wide1971} together with a Markov chain Monte Carlo 
(MCMC) procedure \citep{Gilks1996} to obtain the uncertainties. The geometrical and physical 
parameters of the system are calculated as described in Section~\ref{modelling}. 
These parameters were used as fixed inputs for the WD code and only the times of individual 
primary eclipses are left as free parameters. The median values and the 
1-$\sigma$ uncertainties obtained from the marginal distribution of the fitted instants 
of minimum were adopted as the best values for location and error of each timing. Using 
the expression $T_{\rm min} = T_0 + E \times P$, where $T_{\rm min}$ are 
the predicted times of primary minimum, $T_0$ is a fiducial epoch, $E$ is the cycle 
count from $T_0$, and $P$ is the binary orbital period, we obtained the following ephemeris,
\begin{equation}
T_{\rm min} = {\rm TDB}\,2454274.2087(1) + 0.110374230(2) \times E.
\end{equation}

\subsection{Radial velocity solution}
\label{sec:rvsolution}
\begin{figure}
 \resizebox{\hsize}{!}{\includegraphics{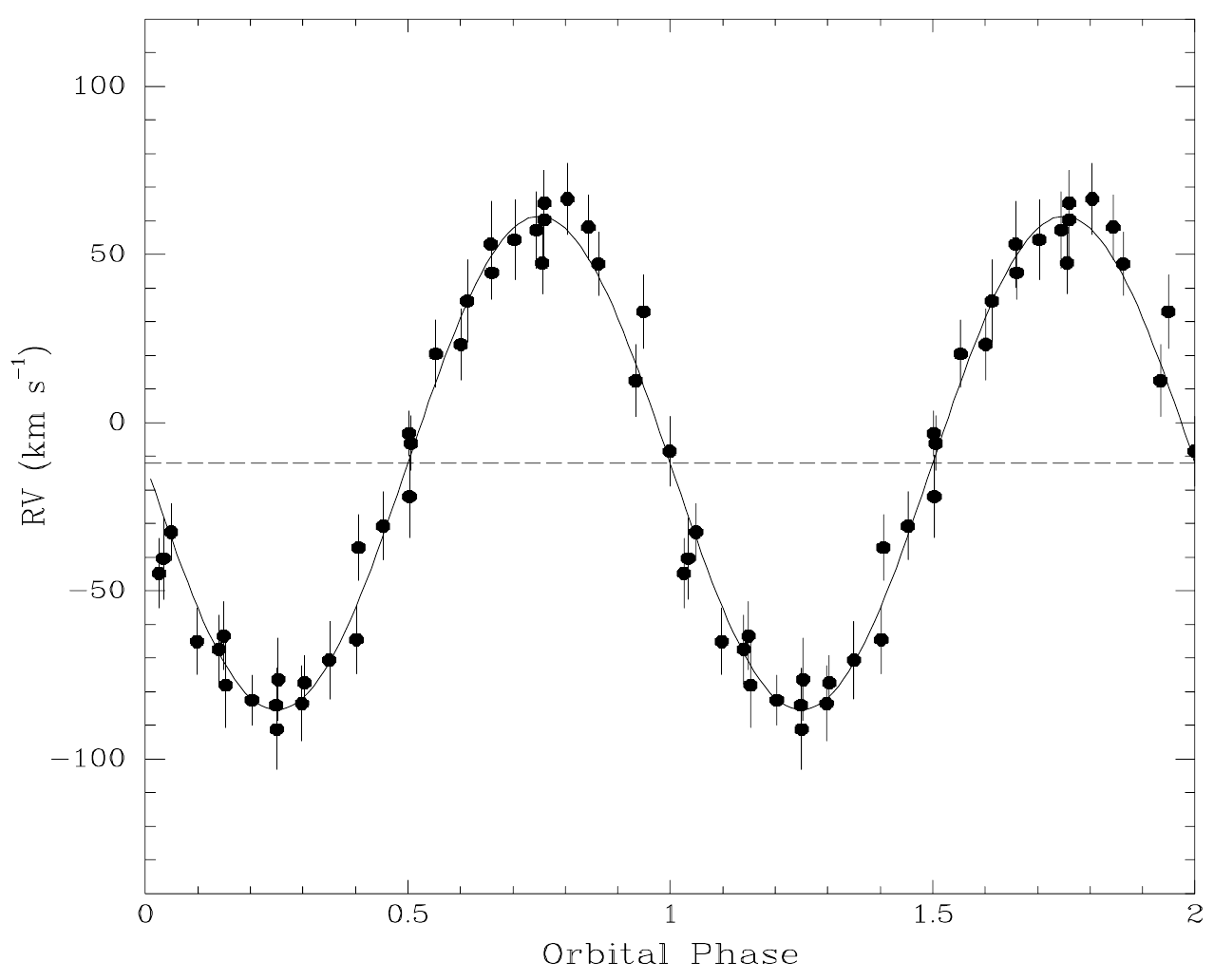}}
 \caption{Radial velocity of the prominent lines in the spectra of NSVS\,1425.
 The phases are calculated using the ephemeris presented 
          in Section~\ref{ephemeris}.}
 \label{fig:RV}
\end{figure}

The radial velocities were obtained using the task {\tt FXCOR} in {\tt IRAF}. Initially,
a combination of all 36 spectra was used as a template for the cross correlation with the
individual spectra. Regions around H$\beta$, H$\gamma$, H$\delta$, 
H$\epsilon$, and HeII\,$\lambda$4686 were selected (see Figure~\ref{fig:spectrum}) to improve the signal/noise ratio of the correlation procedure. The resulting radial velocity 
solution was used to Doppler shift all individual spectra to the orbital rest frame. 
A better quality template is then produced from these rest-frame spectra. The procedure was 
iterated a number of times until the radial velocity solution converged. Table~\ref{table:RV} 
lists individual radial velocities and Figure~\ref{fig:RV} shows the radial velocity 
curve folded on the orbital phase together with the best solution for a circular orbit.
The modelling provides a semi-amplitude $K_1=73.4\pm2.0$~km~s$^{-1}$ and a systemic 
velocity $\gamma = -12.1\pm1.5$~km~s$^{-1}$.

\begin{table}
\begin{center}
\begin{scriptsize}
\caption{Spectroscopic observations and radial velocities.}          
\label{table:RV}                           
\begin{tabular}{l r r r r}       
\hline\hline                
UT Date& BJD(TDB)   & t$_{\rm exp}$ & V~~~~~~~   & Orbital \\    
       & (2450000+) &    (s)        & (km\,s$^{-1}$)~~~ & Phase \\
\hline                     
   2010 Set 01 & 5441.48256 & 900 &    2.29$\pm$12.29& 0.60  \\
   2010 Set 01 & 5441.49251 & 600 &  -8.05$\pm$11.86 & 0.69   \\
   2010 Set 01 & 5441.50371 & 900 & 66.46$\pm$10.50  & 0.79   \\
   2010 Set 01 & 5441.51799 & 900 & 12.46$\pm$10.60  & 0.92  \\
   2010 Set 01 & 5441.52904 & 900 & -40.38$\pm$12.10 & 0.02  \\
   2010 Set 01 & 5441.54078 & 900 & -67.43$\pm$10.21 & 0.13  \\
   2010 Set 01 & 5441.55825 & 900 & -83.52$\pm$11.16 & 0.29  \\
   2010 Set 01 & 5441.56970 & 900 & -64.55$\pm$10.12 & 0.39  \\
   2010 Set 01 & 5441.58084 & 900 & -21.99$\pm$12.32 & 0.49  \\
   2010 Set 01 & 5441.59792 & 900 &  53.03$\pm$12.84 & 0.65  \\
   2010 Set 01 & 5441.60920 & 900 &  60.29$\pm$10.98 & 0.75  \\
   2010 Set 01 & 5441.62057 & 900 &  47.18$\pm$9.33  & 0.85  \\
   2010 Set 01 & 5441.63857 & 900 & -44.83$\pm$10.33 & 0.02  \\  
   2010 Set 01 & 5441.65259 & 900 & -78.09$\pm$12.25 & 0.14  \\  
   2010 Set 01 & 5441.66364 & 900 & -76.39$\pm$12.18 & 0.24  \\  
   2010 Set 02 & 5442.42478 & 900 & -63.45$\pm$10.09 & 0.14  \\   
   2010 Set 02 & 5442.43581 & 900 & -84.04$\pm$10.91 & 0.24  \\   
   2010 Set 02 & 5442.45306 & 900 & -37.19$\pm$9.84  & 0.40  \\   
   2010 Set 02 & 5442.46413 & 900 & -6.20$\pm$8.10   & 0.50  \\   
   2010 Set 02 & 5442.48116 & 900 & 44.56$\pm$8.03   & 0.65  \\   
   2010 Set 02 & 5442.49223 & 900 & 65.25$\pm$9.62   & 0.75  \\   
   2010 Set 02 & 5442.51304 & 900 & 32.99$\pm$10.88  & 0.94  \\   
   2010 Set 02 & 5442.52404 & 900 &-32.57$\pm$8.55   & 0.04  \\   
   2010 Set 02 & 5442.54110 & 900 &-82.53$\pm$7.39   & 0.19  \\ 
   2010 Set 02 & 5442.55209 & 900 &-77.39$\pm$8.10   & 0.29  \\
   2010 Set 02 & 5442.56870 & 900 &-30.80$\pm$10.13  & 0.44  \\
   2010 Set 02 & 5442.57970 & 900 & 20.41$\pm$10.04  & 0.54  \\   
   2010 Set 02 & 5442.60086 & 900 & 57.22$\pm$11.46  & 0.74  \\ 
   2010 Set 02 & 5442.61186 & 900 & 58.08$\pm$9.66   & 0.83  \\   
   2010 Set 02 & 5442.62893 & 900 & -8.53$\pm$10.40  & 0.98 \\  
   2010 Set 02 & 5442.63992 & 900 &-65.09$\pm$9.83   & 0.09 \\  
   2010 Set 02 & 5442.65664 & 900 &-91.21$\pm$10.98  & 0.24 \\
   2010 Set 02 & 5442.66763 & 900 &-70.64$\pm$11.55  & 0.34 \\
   2010 Set 02 & 5442.68442 & 900 & -3.21$\pm$6.66   & 0.49 \\   
   2010 Set 02 & 5442.69542 & 900 & 23.22$\pm$10.57  & 0.59 \\   
   2010 Set 02 & 5442.71246 & 900 & 47.40$\pm$9.12   & 0.74 \\   
\hline                                
\end{tabular}
\end{scriptsize}
\end{center}
\end{table}

\subsection{Atmospheric parameters}\label{spectra_fitting}

The atmospheric parameters of the sdB star can be determined using the Balmer and helium 
lines in the blue range of the spectrum. The spectra obtained in a 0.1 phase interval
centred in the secondary eclipse, i.e., 0.45--0.55, were used to minimise the contribution of the 
reflection effect. Using $\chi^2$ as the figure of merit, the combined spectrum was matched
to a grid of synthetic spectra retrieved from the web page 
of TheoSSA\footnote{http://dc.g-vo.org/theossa}. The synthetic 
spectra were calculated by the T\"ubingen non-local thermodynamic equilibrium Model-Atmosphere Package\footnote{http://astro.uni-tuebingen.de/$\sim$rauch/TMAP/TMAP.html} (TMAP). Two 
different metallicities were used: Model A with zero metallicity; and Model B with the metallicity adopted by \citet{2011A&A...531L...7K}. The grid is composed by: 26 values of effective temperatures, 30000 K $\leq$ T $\leq$ 43000 K with 500 K steps; 16 surface gravities, $5.2 \leq \log g \leq 6.0$ with 0.05 dex steps; and 10 helium abundances, 0.001 $\leq$ $n(\rm He)$/$n(\rm H)$ $\leq$ 0.01 with 0.001 dex steps. All synthetic spectra were convolved with the projected rotational velocity $v\sin\,i = 73.4$ km s$^{-1}$ and with an instrumental profile of FWHM 1.8~\AA.

\begin{figure}
 \centering
 \resizebox{\hsize}{!}{\includegraphics{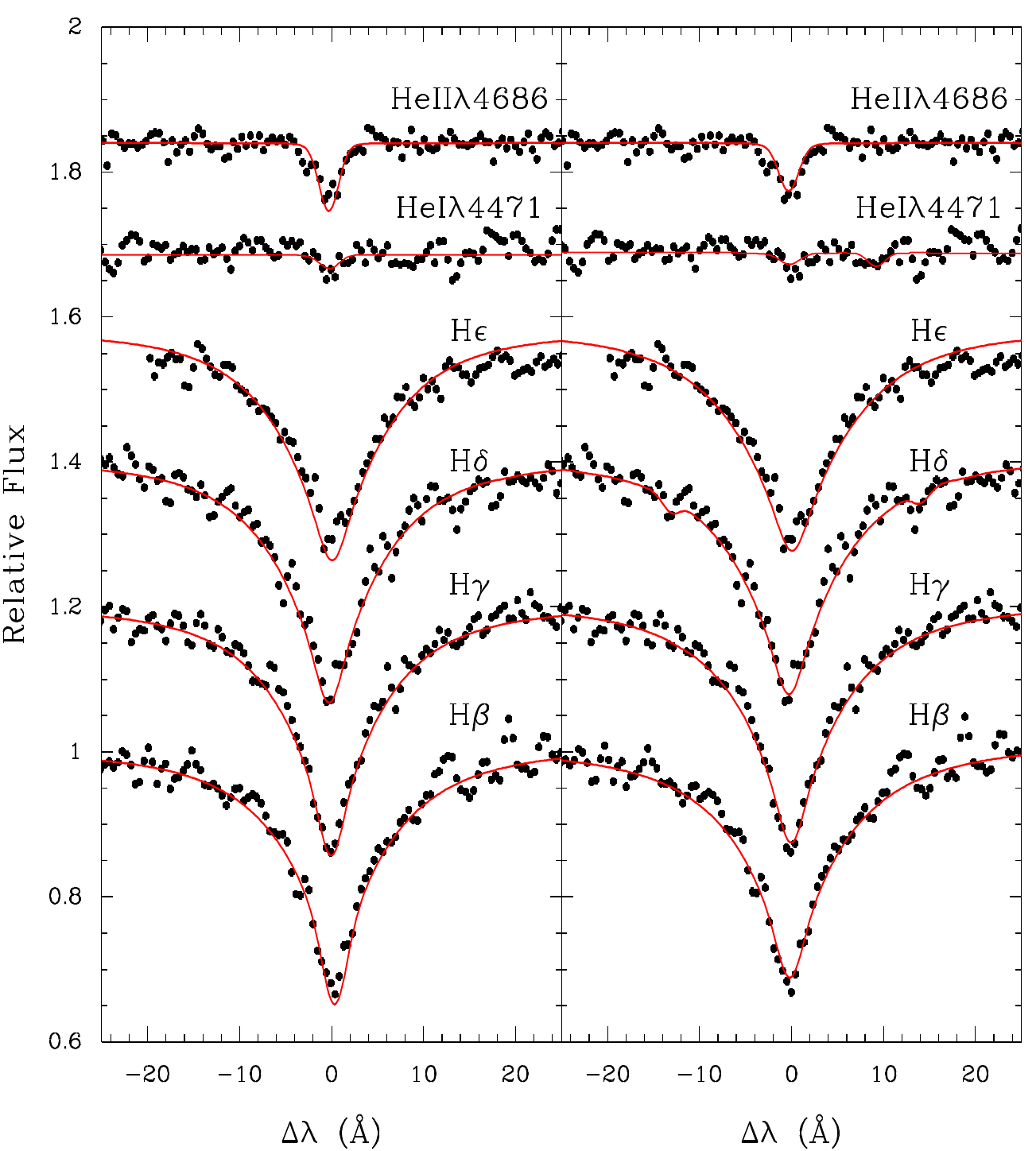}}
 \caption{The best fits to the Balmer and helium lines used to derive effective 
  temperature, surface gravity, and helium abundance. The observed spectra   
  are presented using black dots and the red lines represent the best-fit 
  synthetic spectra. The left and right panels correspond to models 
  with zero metallicity and the metallicity from \citet{2011A&A...531L...7K}, respectively.}
\label{fig:fitspectrum}
\end{figure}

The synthetic Balmer (H$\beta$ to H$\epsilon$) and helium (HeI\,$\lambda 4471$ and 
HeII\,${\lambda4686}$) lines were used in the fitting procedure to determine the effective 
temperature, surface gravity, and He abundance. The best fit yields $T = 38000\pm500$~K, 
$\log g = 5.2\pm0.05$, and $n(\rm He)$/$n(\rm H) = 0.003\pm0.001$ for the Model A, and
$T = 40000\pm500$~K, $\log g = 5.50\pm0.05$, and $n(\rm He)$/$n(\rm H) = 0.003\pm0.001$ for Model B. Figure~\ref{fig:fitspectrum} shows the observed spectra together with the best synthetic spectra for both models. The $\chi^2$ of model B is $\sim$5 per cent better than that of model A.

\subsection{Simultaneous modelling of light curves and radial velocities}
\label{modelling}

In order to obtain the geometrical and physical parameters for NSVS\,1425,
we simultaneously analysed the U, B, V, R$_{\rm C}$, I$_{\rm C}$, J, and H light 
curves and the radial velocity curve using the WD code. The original WD code uses a 
differential correction method for improving an initial solution. This method works 
well if the initial parameter values are close to those corresponding to the optimal 
solution. However, if they are close to a local minimum, the differential correction 
procedure may fail to find the best solution. To solve this problem, the WD 
code was used as a ``function"~to be optimised by the genetic algorithm  {\tt PIKAIA} 
\citep{char1995}, which is adequate to search for a global minimum in a model 
involving a large set of parameters.

To examine the marginal distribution of probability of the parameters and to 
establish realistic uncertainties, we used the solution obtained by {\tt PIKAIA} 
as an input to an MCMC procedure.

Due to the large number of parameters to be fitted, it is important to constrain them using 
theoretical and spectroscopic information. From the spectroscopic analysis, we adopted 
the effective temperature of the primary star as an initial value. We can also constrain
the mass ratio, $q$, using the mass function (Eq. \ref{eq:mass_function}) and assuming that the mass of the sdB star, $M_1$, is in the range $0.1-0.8~\rm M_{\odot}$ \citep{Driebe1998, Han2003}, and that the radial velocity semi-amplitude, $K_1$,  is 73.4~km s$^{-1}$.

\begin{equation}
 \frac{M_1\times (q\sin\,i)^3}{(1+q)^2} = 1.0361\times10^{-7}(1-e^2)^{3/2}K_1^3P.
  \label{eq:mass_function}
\end{equation}

As the components are in close orbit, the timescales of synchronisation and 
circularisation are much shorter than the helium burning lifetime \citep{Zahn1977}. 
Thus, the orbit can be considered circular ($e$ = 0) and the rotation of the components 
synchronised with the orbit.  Finally, adopting the range of orbital inclinations for 
eclipsing binaries, $75^{\circ}<i<90^{\circ}$, we obtained $0.21 < q < 0.45$ for the mass 
ratio range. 

Mode 2 of the WD code, which sets no constraints on the Roche configuration, was used. 
The luminosity of the secondary star was computed assuming stellar atmosphere radiation. 
Linear limb darkening coefficients, $x_i$, were used for both stars. Regarding the sdB star, 
we used the coefficients calculated by \citet*{Diaz1995} and \citet*{Claret1995} for a 
star with effective temperature $T = 40000$~K and surface gravity $\log g = 5.0$. These are the closest values to those of the hot component in NSVS\,1425 for which limb darkening coefficients have been published in literature. On the other hand, the linear limb darkening coefficients of the cool star were left as free  parameters, since the proximity of the hot star can significantly change these coefficients with respect to those of a single star. As the sdB star has a radiative envelope, its gravity darkening exponent, $\beta_1$, and its bolometric albedo for reflective heating and re-radiation, $A_1$, were set to unity \citep{Rafert1980}.  
The gravity darkening exponent of the secondary component, $\beta_2$, was fixed at 0.3,
which is appropriate for convective stars \citep{Lucy1967}. As shown by \citet{For2010}, 
\citet{Kilkenny1998}, and \citet{Drechsel2001}, the albedo of the secondary star, $A_2$, 
can assume physically unrealistic values $A_2 > 1$, especially at longer wavelengths 
where the reflected-reradiated light is more intense. For this reason, it was decided to 
perform two modellings: in Model 1, we adopt a constant (but free parameter) secondary albedo for all photometric bands; in Model 2, we consider variable and independent albedos for all photometric bands.  
 
In both cases the remaining fitted parameters consist of: the mass ratio, $q = M_2/M_1$; 
the orbital inclination, $i$; the separation between the components, $a$; the Roche potentials, 
$\Omega_1$ and $\Omega_2$; and the effective temperatures of the two stars, $T_1$ and $T_2$. 

\begin{figure}
 \resizebox{\hsize}{!}{\includegraphics{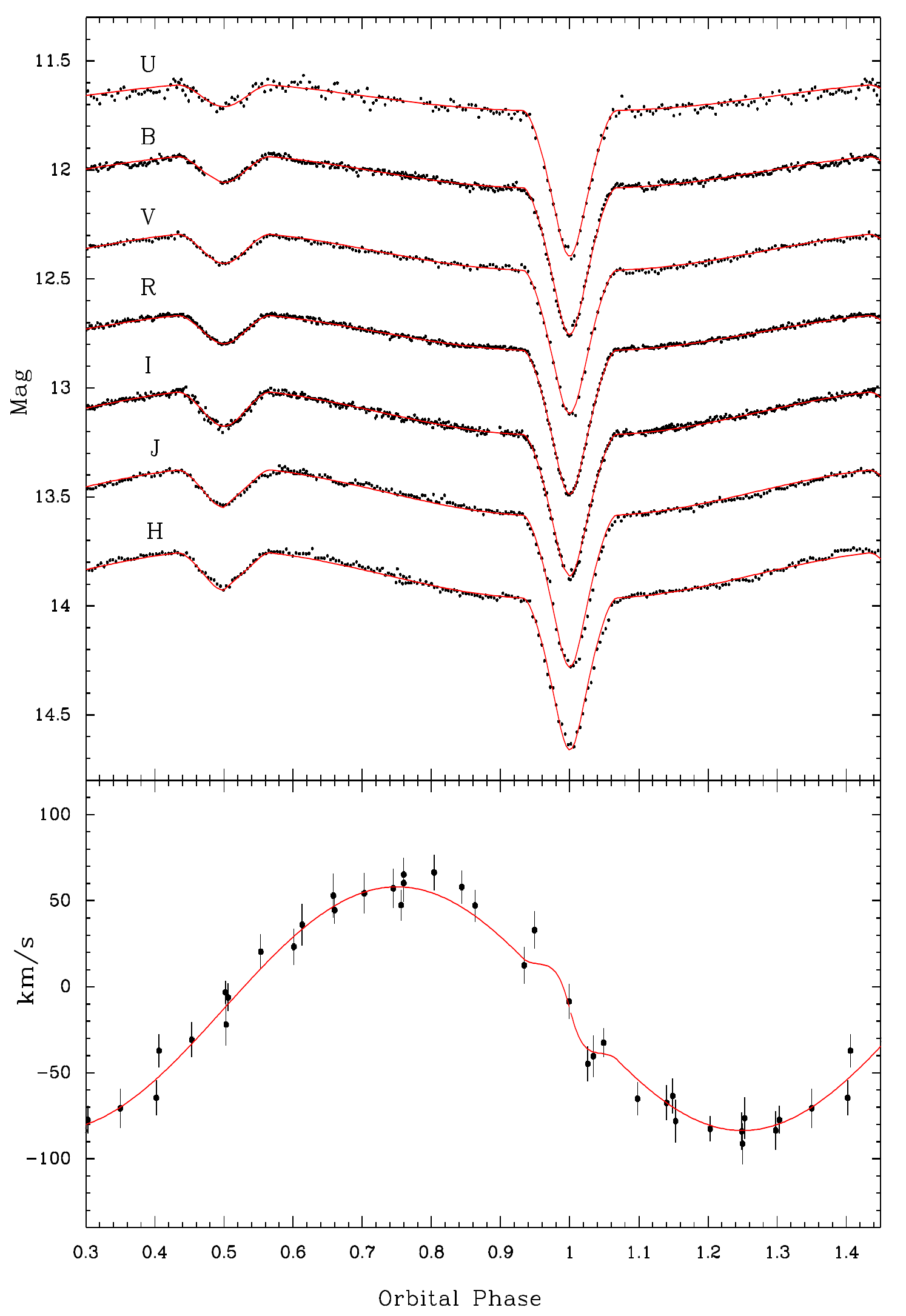}}
 \caption{The best simultaneous fit for the U, B, V, R$_{\rm C}$, I$_{\rm C}$, J, H 
light curves and radial velocity data using the WD code. The light curves are the same as those shown in Figure \ref{light_curve} displaced vertically for better visualisation.}
 \label{lc_rv}
\end{figure}

In order to optimise the computational time, all light curves were binned with 160 seconds 
time resolution and the error of the bin average outside of the eclipses was assumed as the 
uncertainty. To test the goodness of fit, we use the reduced $\chi^2$ defined as

\begin{equation}
\centering
\chi^2= \frac{1}{n}\sum_i^n\left(\frac{O_i-C_i}{\sigma_i}\right)^2,
\end{equation}

\noindent where $O_i$ are the observed points, $C_i$ are the corresponding model, $\sigma_i$ are the uncertainties at each point, and $n$ is the number of points. Figure~\ref{lc_rv} 
shows the best fit together with the light and radial velocity curves while 
Table~\ref{sytem:parameters} lists the fitted and fixed parameters.

\begin{table}
\small
\begin{center}
\caption{System parameters of the best fit for photometric light curves in U,
         B, V, R$_{\rm C}$, I$_{\rm C}$, J and H bands and radial velocity data for 
         NSVS\,1425.}         
\label{sytem:parameters}    
\begin{tabular}{l r r}       
\hline
\hline
 Parameter          & Model 1        & Model 2   \\
\hline
Fitted Parameters & \\
\hline
$q=(M_2/M_1)$       & $0.28\pm0.013$  &  $0.26\pm0.012$  \\
$i~(^{\circ}$)      & $82.5\pm0.4$   &  $82.5\pm0.3$     \\
$\Omega^a_1 $       & $4.58\pm0.13$  &  $4.55\pm0.11$    \\
$\Omega^a_2$        & $2.80\pm0.15$  &  $2.69\pm0.12$    \\
$T_1$(K)            & $42300\pm400$  &  $42000\pm500$    \\
$T_2$(K)            & $2400\pm600$   &  $2550\pm550$     \\
$a^b$ (R$_{\odot}$) & $0.74\pm0.04$  &  $0.80\pm0.04$    \\ 
$A^c_2$(U)          & $1.50\pm0.11$  &  $2.0\pm0.15$      \\
$A^c_2$(B)          & $1.50\pm0.11$  &  $1.35\pm0.13$    \\
$A^c_2$(V)          & $1.50\pm0.11$  &  $1.20\pm0.12$    \\
$A^c_2$(R$_{\rm C}$)& $1.50\pm0.11$  &  $1.05\pm0.09$    \\
$A^c_2$(I$_{\rm C}$)& $1.50\pm0.11$  &  $1.3\pm0.12$      \\ 
$A^c_2$(J)          & $1.50\pm0.11$  &  $0.95\pm0.14$    \\ 
$A^c_2$(H)          & $1.50\pm0.11$  &  $1.10\pm0.15$    \\ 
$x_2$(U)            & $0.64\pm0.04$  &  $0.68\pm0.04$    \\
$x_2$(B)            & $0.69\pm0.04$  &  $0.74\pm0.05$    \\
$x_2$(V)            & $0.78\pm0.03$  &  $0.80\pm0.04$    \\
$x_2$(R$_{\rm C}$)  & $0.83\pm0.02$  &  $0.87\pm0.02$    \\
$x_2$(I$_{\rm C}$)  & $0.90\pm0.03$  &  $0.92\pm0.03$    \\
$x_2$(J)            & $0.93\pm0.04$  &  $0.95\pm0.04$    \\
$x_2$(H)            & $0.98\pm0.05$  &  $0.99\pm0.05$    \\
\hline
Roche radii$^h$  &  & \\
\hline
$r_1$ (pole)   & 0.231$\pm$0.006 &   0.233$\pm$0.005  \\
$r_1$ (side)   & 0.233$\pm$0.006 &   0.235$\pm$0.005  \\
$r_1$ (point)  & 0.235$\pm$0.007 &   0.236$\pm$0.006  \\
$r_1$ (back)   & 0.234$\pm$0.007 &   0.236$\pm$0.006  \\
$r_2$ (pole)   & 0.180$\pm$0.016 &   0.194$\pm$0.014  \\
$r_2$ (side)   & 0.182$\pm$0.016 &   0.198$\pm$0.016  \\ 
$r_2$ (point)  & 0.191$\pm$0.019 &   0.210$\pm$0.019  \\ 
$r_2$ (back)   & 0.189$\pm$0.019 &   0.207$\pm$0.018  \\
\hline
 Fixed Parameters  & \\
\hline
$A^c_1$            & 1.0   & 1.0   \\
$\beta^d_1$        & 1.0   & 1.0   \\
$\beta^d_2$        & 0.3   & 0.3   \\
$x^e_1$(U)         & 0.242 & 0.242 \\
$x^e_1$(B)         & 0.233 & 0.233 \\
$x^e_1$(V)         & 0.209 & 0.209 \\
$x^f_1$(R$_{\rm C}$)&0.176 & 0.176 \\
$x^f_1$(I$_{\rm C}$)&0.147 & 0.147 \\
$x^f_1$(J)         & 0.112 & 0.112 \\
$x^f_1$(H)         & 0.095 & 0.095 \\
\hline
Goodness of fit  \\
\hline                 
$\chi^2$           & 2.1   & 1.2 \\
\hline                   
\end{tabular}
\end{center}
$^a$ Roche surface potential; \\
$^b$ Components separation;\\
$^c$ Bolometric albedo;\\
$^d$ Gravity darkening exponent;\\
$^e$ Linear limb darkening coefficient from \citet{Diaz1995};\\
$^f$ Linear limb darkening coefficient from \citet{Claret1995};\\
$^h$ In units of orbital separation.
\end{table}

The values found for almost all parameters in the two models are consistent.
However, the $\chi^2$ for Model 2 is $\sim$40 per cent better than for Model 1. 

\subsection{Fundamental Parameters}

Physical and geometrical parameters such as masses, radii, and separation between
the two components of the system can be 
derived from the solutions obtained in the previous sections. Substituting the values of 
the orbital period ($P_{\rm orb} = 0.110374230$ d), semi-amplitude of the radial velocity 
($K_1 = 73.4$~km s$^{-1}$), mass ratio ($q = 0.260$), and inclination ($i = 82\fdg5$) 
in Eq. \ref{eq:mass_function}, $M_1 = 0.419 \pm 0.070~\rm M_{\odot}$ is obtained for the 
primary mass. The primary mass and the mass ratio are used to derive the secondary mass, 
$M_2 = 0.109 \pm 0.023~\rm M_{\odot}$. Using Kepler's Third Law, one can obtain the orbital 
separation, $a = 0.80\pm 0.04~\rm R_{\odot}$, from which the absolute radii follow, 
$R_1 = 0.188 \pm 0.010~\rm R_{\odot}$ and $R_2 = 0.162 \pm 0.008~\rm R_{\odot}$.
In Table~\ref{fundamental:parameters} we show the fundamental parameters for 
NSVS\,1425 derived from Model 1 and Model 2.

\begin{table}
\begin{center}
\caption{Fundamental parameters for NSVS\,1425.}            
\label{fundamental:parameters}      
\begin{tabular}{l r r}        
\hline
\hline                 
 Parameter & Model 1 & Model 2 \\     
\hline
$M_1$ (M$_{\odot}$)     &  0.346$\pm$0.079 & 0.419$\pm$0.070 \\
$M_2$ (M$_{\odot}$)     &  0.097$\pm$0.028 & 0.109$\pm$0.023 \\
$R_1$ (R$_{\odot}$)     &  0.173$\pm$0.010 & 0.188$\pm$0.010 \\
$R_2$ (R$_{\odot}$)     &  0.137$\pm$0.008 & 0.162$\pm$0.008 \\
$T_1$ (K)               &  42300$\pm$500   & 42000$\pm$400   \\
$T_2$ (K)               &   2400$\pm$500   &  2550$\pm$500   \\
$\log g_1$              &   5.50$\pm$0.14  &  5.51$\pm$0.11  \\
$\log g_2$              &   5.15$\pm$0.16  &  5.05$\pm$0.13  \\
$a$~(R$_{\odot}$)       &   0.74$\pm$0.04  &  0.80$\pm$0.04  \\
\hline
\end{tabular}
\end{center}
\end{table}

\subsection{Rossiter-McLaughlin effect}
 
An interesting spectroscopic signature in eclipsing binary systems is the 
Rossiter-McLaughlin (RM) effect \citep{Rossiter1924, McLaughlin1924}. This 
effect occurs when the eclipsed object rotates. There is evidence of this effect 
on the radial velocity curve in the phase interval $0.95-1.05$ (Figure~\ref{lc_rv}). 
Unfortunately, our data are not enough to model the RM 
effect to derive the alignment rotational parameters of the NSVS\,1425 stars. 
However, our observed points are consistent with the predicted RM effect for aligned 
rotating stars obtained by the WD code with the parameters shown in 
Table~\ref{sytem:parameters}.

\section{Discussion}

\subsection{Characteristics of the primary star}

Of all HW Vir systems, the primary of NSVS\,1425 has the second highest 
temperature (see Table~\ref{tab:known}), consistent with the prominent HeII\,$\lambda$4686 
line (see Figure~\ref{fig:spectrum}). 
Accordingly, we suggest that the primary of NSVS\,1425 is an sdBO star which means
this system is very similar to AA Dor and, hence, in a rare evolutionary stage 
\citep{Heber2009}.

Comparing the values of $\log g$ derived from the simultaneous fit to photometric and 
spectroscopic data (Table~\ref{fundamental:parameters}) with those obtained 
from the modelling of the spectral lines of the primary star (Section~\ref{spectra_fitting}), 
it is clear that the model with metallicity equal to that adopted by \citet{2011A&A...531L...7K} 
provides consistent results, whereas the model with zero metallicity has a discrepancy. 
We noticed that the same kind of discrepancy 
had been found by \citet{2000A&A.356.665R} in the analysis of the primary in AA Dor. 
\citet{2000A&A.356.665R} obtained $\log g = 5.21$ from spectroscopic data, whereas 
\citet{1996MNRAS.279.1380H} had derived $\log g  = 5.53$ from photometric data modelling. 
This discrepancy was solved by \citet{2011A&A...531L...7K} with the improvement of the 
Stark broadening modelling, the minimisation of the reflection effect, and adoption 
of metal-line blanketing.

\subsection{Evolution}

\citet{Han2003}, from a detailed binary population synthesis study, presented three 
possible channels for forming sdB stars: 

\begin{enumerate}

 \item \textit{One or two phases of common envelope (CE) evolution};

 \item \textit{One or two stable Roche lobe overflows};

 \item \textit{A merger of two He-core white dwarfs}.

\end{enumerate}

\citet{Driebe1998} and \citet{Heber2003} suggest another scenario to form an sdB star with  
low mass in binaries called post-RGB. This scenario is similar to the channel (i) proposed by 
\citet{Han2003}, except that the resultant sdB in the first CE phase has insufficient mass in its
core to ignite helium. However, it will evolve as a helium star through the sdB 
star region to form a helium core white dwarf.

\begin{figure}
 \resizebox{\hsize}{!}{\includegraphics[angle=-90]{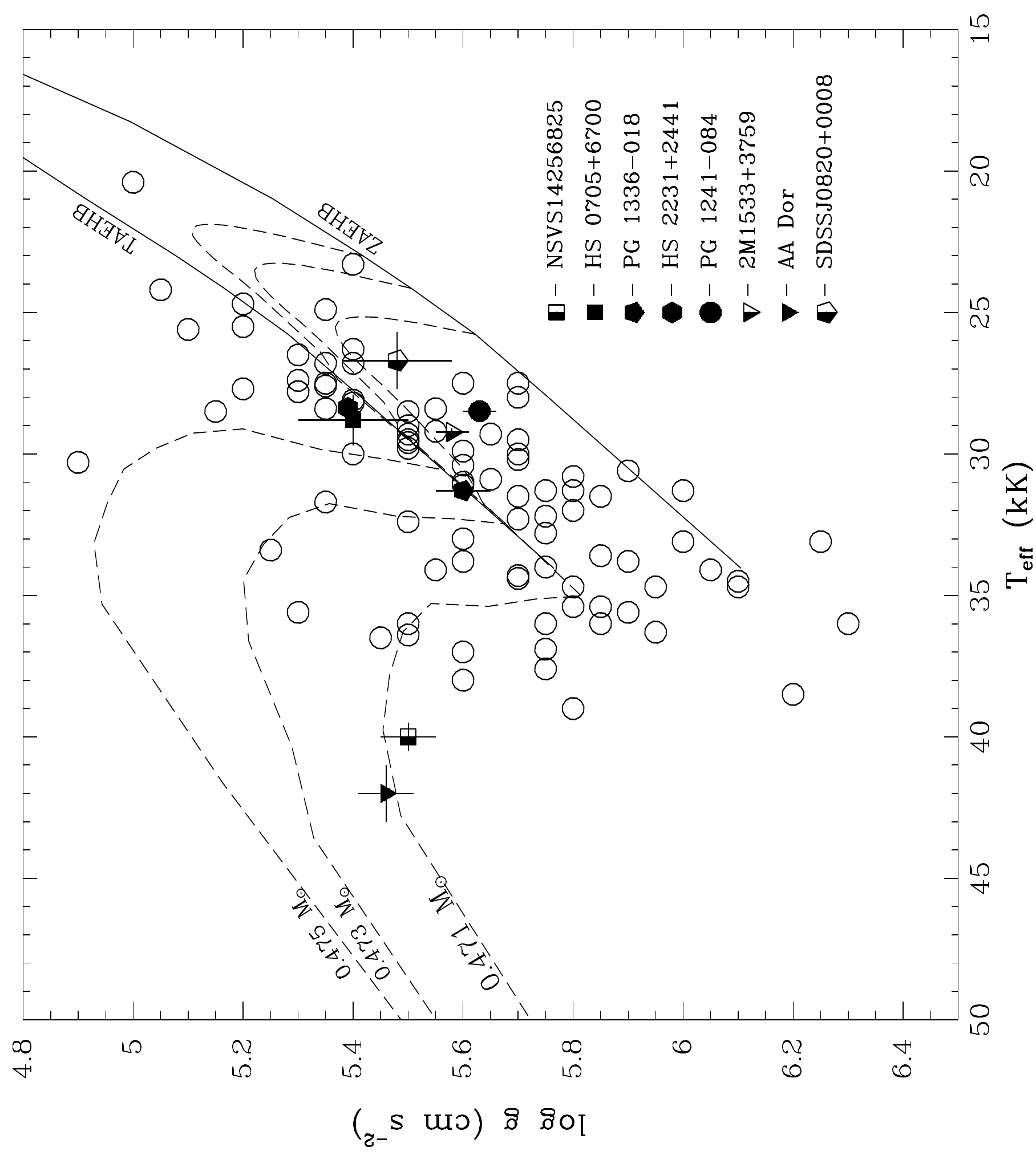}}
 \caption{Position on the ($T_{\rm eff}$, $\log g$) diagram of the hot component of 
 NSVS\,1425 compared with other sdB stars (see Table~\ref{tab:known}).
Isolated sdB and sdOB stars presented in \citet{Edelmann2003} are shown with open circles. 
Dashed lines represent evolutionary tracks for different masses in the post-EHB evolution 
\citep{Dorman1993}. The zero age extreme horizontal branch (ZAEHB) and terminal age extreme 
horizontal branch (TAEHB) are represented by solid lines.}
 \label{txlogg}
\end{figure}

In Figure~\ref{txlogg} we compare the position of the primary component of NSVS\,1425 
on the ($T_{\rm eff}$, $\log g$) diagram with other sdB stars in short-period binary 
systems (see Table~\ref{tab:known}). We also show a sample of single sdB and sdOB stars 
analysed by \citet{Edelmann2003}. In the same graph we display evolutionary tracks for different 
masses in the post-EHB calculated by \citet*{Dorman1993} for single star evolution.
As can be seen, the position of the NSVS\,1425 primary is
close to that of the AA Dor primary. The evolutionary track for single star evolution 
would only marginally explain the mass obtained for the sdB star 
in NSVS\,1425. Out of the possible channels to form an sdB in binaries, we suggest that the 
channel (i) is probably the evolutionary scenario for NSVS\,1425. 

\section{Conclusion}
We present a photometric and spectroscopic analysis of the NSVS\,1425 system. 
With a short orbital period, $P=0.110374230(2)$ d, this binary shows both primary and 
secondary eclipses and a prominent reflection effect. 
From the spectroscopic analysis we obtain $73.4 \pm 2.0~\rm km~s^{-1}$  for the 
semi-amplitude of the radial velocity and $-12.1 \pm 1.5~\rm km~s^{-1}$  
for the systemic velocity. The atmospheric parameters of the primary component (sdOB star), 
namely, effective temperature,  $T = 40000\pm500$ K, surface gravity, 
$\log g = 5.5\pm 0.05$, and Helium abundance, $n(\rm He)$/$n(\rm H) = 0.003\pm0.001$, 
were calculated matching the observed spectrum to a grid of NLTE synthetic spectra. 

Simultaneously fitting the U, B, V, R$_{\rm C}$, I$_{\rm C}$, J, and H-bands light curves 
and radial velocity curve using the WD code, the geometrical and physical 
parameters of NSVS\,1425 were obtained. These results allow us to derive the absolute 
parameters of the system such as masses and radii of the components.

We compare the position of the sdB star in NSVS\,1425 with other sdB and sdOB stars on the effective temperature versus surface gravity diagram. We describe the 
possible channels to form an sdB star in binaries and conclude that the post-common envelope
development is probably the evolutionary  scenario for NSVS\,1425. The subsequent 
evolution of this system should lead to a cataclysmic variable. After a phase of angular 
momentum loss via gravitational radiation, this system will lie below the period gap of 
the cataclysmic variables. 

\section*{Acknowledgements}
  This study was partially supported by CAPES (LAA and JT), CNPq (CVR: 308005/2009-0), 
  and Fapesp (CVR: 2010/01584-8). The TheoSSA service (http://dc.g-vo.org/theossa) used 
  to retrieve 
  theoretical spectra for this paper was constructed as part of the activities of the 
  German Astrophysical Virtual Observatory. We acknowledge the use of the SIMBAD 
  database, operated at CDS, Strasbourg, France; the NASA's Astrophysics Data System 
  Service; and the NASA's {\it SkyView} facility  (http://skyview.gsfc.nasa.gov) 
  located at NASA Goddard Space Flight Center.

\end{document}